\documentclass[12pt,a4paper]{article}
\usepackage[]{graphicx}

\begin{document}
 \,\,\,\,

\vskip 1.5cm \centerline{\bf CHEMICALLY INDUCED NANOSCALE
JOSEPHSON EFFECTS} \centerline{\bf IN NON-STOICHIOMETRIC
HIGH-T$_{\bf C}$ SUPERCONDUCTORS} \vspace{15mm}

\centerline{\bf SERGEI SERGEENKOV$^{a,b}$} \vspace{8mm}

\centerline{$^{a}$Centro de F\'\i sica das Interac\c c\~oes
Fundamentais,} \centerline{Instituto Superior T\'ecnico, Lisboa,
Portugal} \centerline{$^{b}$Bogoliubov Laboratory of Theoretical
Physics,} \centerline{Joint Institute for Nuclear Research, Dubna,
Russia} \vspace{20mm}

\leftline{\bf 1. INTRODUCTION} \vspace{5mm}

Both granular superconductors and artificially prepared arrays of
Josephson junctions (JJAs) proved useful in studying the numerous
quantum (charging) effects in these interesting systems, including
blockade of Cooper pair tunneling~\cite{1}, Bloch
oscillations~\cite{2}, propagation of quantum ballistic
vortices~\cite{3}, spin-tunneling related effects using specially
designed $SFS$-type junctions~\cite{4,5}, novel Coulomb effects in
$SINIS$-type nanoscale junctions~\cite{6}, dynamical AC
reentrance~\cite{7} and geometric quantization phenomena~\cite{8}
in 2D JJAs (see, e.g., Ref.~\cite{9} for the recent review on
charge and spin effects in mesoscopic 2D Josephson junctions).

More recently, it was realized that JJAs can be also used as
quantum channels to transfer quantum information between distant
sites~\cite{10,11,12} through the implementation of the so-called
superconducting qubits which take advantage of both charge and
phase degrees of freedom (see, e.g., Ref.~\cite{13} for a review
on quantum-state engineering with Josephson-junction devices).

Recent imaging of the granular structure in underdoped
$Bi_2Sr_2CaCu_2O_{8+\delta}$ crystals~\cite{14}, revealed an
apparent segregation of its electronic structure into
superconducting domains (of the order of a few nanometers) located
in an electronically distinct background. In particular, it was
found that at low levels of hole doping ($\delta <0.2$), the holes
become concentrated at certain hole-rich domains. Tunneling
between such domains leads to intrinsic granular superconductivity
(GS) in high-$T_c$ superconductors (HTS). Probably one of the
first examples of GS was observed in $YBa_2Cu_3O_{7-\delta }$
single crystals in the form of the so-called "fishtail" anomaly of
magnetization~\cite{15}. The granular behavior has been related to
the 2D clusters of oxygen defects forming twin boundaries (TBs) or
dislocation walls within $CuO$ plane that restrict supercurrent
flow and allow excess flux to enter the crystal. Indeed, there are
serious arguments to consider the TB in HTS as insulating regions
of the Josephson SIS-type structure. An average distance between
boundaries is essentially less than the grain size. In particular,
the networks of localized grain boundary dislocations with the
spacing ranged from $10 nm$ to $100 nm$ have been
observed~\cite{15} which produce effectively continuous normal or
insulating barriers at the grain boundaries. It was also verified
that the processes of the oxygen ordering in HTS leads to the
continuous change of the lattice period along TB with the change
of the oxygen content. Besides, a destruction of bulk
superconductivity in these non-stoichiometric materials with
increasing the oxygen deficiency parameter $\delta $ was found to
follow a classical percolation theory~\cite{16}.

In addition to their importance for understanding the underlying
microscopic mechanisms governing HTS materials, the above
experiments can provide rather versatile tools for designing
chemically-controlled atomic scale Josephson junctions (JJs) and
their arrays (JJAs) with pre-selected properties needed for
manufacturing the modern quantum devices~\cite{10,17}. Moreover,
as we shall see below, GS based phenomena can shed some light on
the origin and evolution of the so-called paramagnetic Meissner
effect (PME) which manifests itself both in high-$T_c$ and
conventional superconductors~\cite{18,19} and is usually
associated with the presence of $\pi$-junctions and/or
unconventional ($d$-wave) pairing symmetry.

This Chapter reviews some of the recently suggested novel effects
which are expected to occur in intrinsically granular
non-stoichiometric material modeled by 2D JJAs which are created
by a regular 2D network of twin-boundary (TB) dislocations with
strain fields acting as an insulating barrier between hole-rich
domains in underdoped crystals. In Section 2 we consider
phase-related magnetization effects, including Josephson
chemomagnetism (chemically induced magnetic moment in zero applied
magnetic field) and its influence on a low-field magnetization
(chemically induced PME), and magnetoconcentration effect
(creation of extra oxygen vacancies in applied magnetic field) and
its influence on a high-field magnetization (chemically induced
analog of "fishtail" anomaly). Section 3 addresses charge-related
phenomena which are actually dual to the chemomagnetic effects
described in Section 2. More specifically, we discuss a possible
existence of a non-zero electric polarization (chemomagnetoelectic
effect) and the related change of the charge balance in
intrinsically granular non-stoichiometric material under the
influence of an applied magnetic field. In particular, we predict
an anomalous low-field magnetic behavior of the effective junction
charge and concomitant magnetocapacitance in paramagnetic Meissner
phase and a charge analog of "fishtail"-like anomaly at high
magnetic fields as well as field-dependent weakening of the
chemically-induced Coulomb blockade.

\vspace{8mm} \leftline{\bf 2. CHEMOMAGNETISM AND FISHTAIL ANOMALY} \vspace{5mm}

As is well-known, the presence of a homogeneous chemical potential
$\mu$ through a single JJ leads to the AC Josephson effect with
time dependent phase difference $\partial \phi /\partial t=\mu
/\hbar$. In this paper, we will consider some effects in
dislocation induced JJ caused by a local variation of excess hole
concentration $c({\bf x})$ under the chemical pressure (described
by inhomogeneous chemical potential $\mu ({\bf x})$) equivalent to
presence of the strain field of 2D dislocation array $\epsilon
({\bf x})$ forming this Josephson contact.

To understand how GS manifests itself in non-stoichiometric
crystals, let us invoke an analogy with the previously discussed
dislocation models of twinning-induced superconductivity~\cite{20}
and grain-boundary Josephson junctions~\cite{21}. Recall that
under plastic deformation, grain boundaries (GBs) (which are the
natural sources of weak links in HTS), move rather rapidly via the
movement of the grain boundary dislocations (GBDs) comprising
these GBs. Using the above evidence, in the previous
paper~\cite{21} we studied numerous piezomagnetic effects in
granular superconductors under mechanical loading. At the same
time, observed~\cite{14,15,22,23,24} in HTS single crystals
regular 2D dislocation networks of oxygen depleted regions
(generated by the dissociation of $<110>$ twinning dislocations)
with the size $d_0$ of a few Burgers vectors, forming a triangular
lattice with a spacing $d\ge d_0$ ranging from $10nm$ to $100nm$,
can provide quite a realistic possibility for existence of 2D
Josephson network within $CuO$ plane. Recall furthermore that in a
$d$-wave orthorhombic $YBCO$ crystal TBs are represented by
tetragonal regions (in which all dislocations are equally spaced
by $d_0$ and have the same Burgers vector ${\bf a}$ parallel to
$y$-axis within $CuO$ plane) which produce screened strain
fields~\cite{23} $\epsilon ({\bf x})=\epsilon _0e^{-{\mid{{\bf
x}}\mid}/d_0}$ with ${\mid{{\bf x}}\mid}=\sqrt{x^2+y^2}$.

Though in $YBa_2Cu_3O_{7-\delta }$ the ordinary oxygen diffusion
$D=D_0e^{-U_d/k_BT}$ is extremely slow even near $T_c$ (due to a
rather high value of the activation energy $U_d$ in these
materials, typically $U_d\simeq 1eV$), in underdoped crystals
(with oxygen-induced dislocations) there is a real possibility to
facilitate oxygen transport via the so-called osmotic (pumping)
mechanism~\cite{25,26} which relates a local value of the chemical
potential (chemical pressure) $\mu ({\bf x})=\mu (0)+\nabla \mu
\cdot {\bf x}$ with a local concentration of point defects as
follows $c({\bf x})=e^{-\mu ({\bf x})/k_BT}$. Indeed, when in such
a crystal there exists a nonequilibrium concentration of
vacancies, dislocation is moved for atomic distance $a$ by adding
excess vacancies to the extraplane edge. The produced work is
simply equal to the chemical potential of added vacancies. What is
important, this mechanism allows us to explicitly incorporate the
oxygen deficiency parameter $\delta $ into our model by relating
it to the excess oxygen concentration of vacancies $c_v\equiv
c(0)$ as follows $\delta=1-c_v$. As a result, the chemical
potential of the single vacancy reads $\mu _v\equiv \mu
(0)=-k_BT\log (1-\delta )\simeq k_BT\delta $. Remarkably, the same
osmotic mechanism was used by Gurevich and Pashitskii~\cite{23} to
discuss the modification of oxygen vacancies concentration in the
presence of the TB strain field. In particular, they argue that
the change of $\epsilon ({\bf x})$ under an applied or chemically
induced pressure results in a significant oxygen redistribution
producing a highly inhomogeneous filamentary structure of
oxygen-deficient nonsuperconducting regions along GB~\cite{24}
(for underdoped superconductors, the vacancies tend to concentrate
in the regions of compressed material). Hence, assuming the
following connection between the variation of mechanical and
chemical properties of planar defects, namely $\mu ({\bf
x})=K\Omega _0\epsilon ({\bf x})$ (where $\Omega _0$ is an
effective atomic volume of the vacancy and $K$ is the bulk elastic
modulus), we can study the properties of TB induced JJs under
intrinsic chemical pressure $\nabla \mu$ (created by the variation
of the oxygen doping parameter $\delta $). More specifically, a
single $SIS$ type junction (comprising a Josephson network) is
formed around TB due to a local depression of the superconducting
order parameter $\Delta ({\bf x})\propto \epsilon({\bf x})$ over
distance $d_0$ producing thus a weak link with (oxygen deficiency
$\delta $ dependent) Josephson coupling $J(\delta )=\epsilon({\bf
x})J_0=J_0(\delta )e^{-{\mid{{\bf x}}\mid}/d_0}$ where $J_0(\delta
)=\epsilon _0J_0=(\mu _v/K\Omega _0 )J_0$ (here $J_0\propto \Delta
_0/R_n$ with $R_n$ being a resistance of the junction). Thus, the
considered here model indeed describes chemically induced GS in
underdoped systems (with $\delta \neq 0$) because, in accordance
with the observations, for stoichiometric situation (when $\delta
\simeq 0$), the Josephson coupling $J(\delta ) \simeq 0$ and the
system loses its explicitly granular signature.

To adequately describe chemomagnetic properties
of an intrinsically granular superconductor, we employ a model of 2D
overdamped Josephson junction array which is based on the well known
Hamiltonian
\begin{equation}
{\cal H}=\sum_{ij}^NJ_{ij}(1-\cos \phi_{ij})+\sum_{ij}^N \frac{q_iq_j}{C_{ij}}
\end{equation}
and introduces a short-range interaction between
$N$ junctions (which are formed around oxygen-rich superconducting
areas with phases $\phi _i(t)$), arranged in a two-dimensional (2D)
lattice with coordinates ${\bf x_i}=(x_i,y_i)$. The areas are
separated by oxygen-poor insulating boundaries (created by TB strain
fields $\epsilon({\bf x}_{ij})$) producing a short-range Josephson
coupling $J_{ij}=J_0(\delta )e^{-{\mid{{\bf x}_{ij}}\mid}/d}$. Thus,
typically for granular superconductors, the Josephson energy of the
array varies exponentially with the distance ${\bf x}_{ij}={\bf
x}_{i}-{\bf x}_{j}$ between neighboring junctions (with $d$ being an
average junction size). As usual, the second term in the rhs of Eq.(1)
accounts for Coulomb effects where $q_i =-2en_i$ is the junction charge
with $n_i$ being the pair number operator. Naturally, the same strain
fields $\epsilon({\bf x}_{ij})$ will be responsible for dielectric properties
of oxygen-depleted regions as well via the $\delta $-dependent capacitance tensor
$C_{ij}(\delta )=C[\epsilon({\bf x}_{ij})]$.

If, in addition to the chemical pressure $\nabla \mu ({\bf
x})=K\Omega _0\nabla \epsilon ({\bf x})$, the network of
superconducting grains is under the influence of an applied
frustrating magnetic field ${\bf B}$, the total phase difference
through the contact reads
\begin{equation}
\phi _{ij}(t)=\phi ^0_{ij}+\frac{\pi w}{\Phi _0} ({\bf x}_{ij}\wedge
{\bf n}_{ij})\cdot {\bf B}+\frac{\nabla \mu \cdot {\bf
x}_{ij}t}{\hbar},
\end{equation}
where $\phi ^0_{ij}$ is the initial phase difference (see below),
${\bf n}_{ij}={\bf X}_{ij}/{\mid{{\bf X}_{ij}}\mid}$ with $ {\bf
X}_{ij}=({\bf x}_{i}+{\bf x}_{j})/2$, and $w=2\lambda _L(T)+l$
with $\lambda _L$ being the London penetration depth of
superconducting area and $l$ an insulator thickness (which, within
the discussed here scenario, is simply equal to the TB
thickness~\cite{26}).

To neglect the influence of the self-field effects in a real
material, the corresponding Josephson penetration length $\lambda
_J=\sqrt{\Phi _0/2\pi \mu _0j_c w}$ must be larger than the junction
size $d$. Here $j_c$ is the critical current density of
superconducting (hole-rich) area. As we shall see below, this
condition is rather well satisfied for HTS single crystals.

Within our scenario, the sheet magnetization
${\bf M}$ of 2D granular superconductor is defined via the average
Josephson energy of the array
\begin{equation}
<{\cal H}>=\int_0^\tau \frac{dt}{\tau}\int \frac{d^2x}{s} {\cal
H}({\bf x},t)
\end{equation}
as follows
\begin{equation}
 {\bf M}({\bf B},\delta )\equiv -\frac{\partial
<{\cal H}>}{\partial {\bf B}},
\end{equation}
where $s=2\pi d^2$ is properly defined normalization area, $\tau$
is a characteristic Josephson time, and we made a usual
substitution $\frac{1}{N}\sum_{ij}A_{ij}(t) \to \frac{1}{s}\int
d^2x A({\bf x},t)$ valid in the long-wavelength
approximation~\cite{27}.

To capture the very essence of the superconducting analog of the
chemomagnetic effect, in what follows we assume for simplicity that a
{\it stoichiometric sample} (with $\delta \simeq 0$) does not possess
any spontaneous magnetization at zero magnetic field (that is
$M(0,0)=0$) and that its Meissner response to a small applied field
$B$ is purely diamagnetic (that is $M(B,0)\simeq -B$). According to
Eq.(4), this condition implies $\phi _{ij}^0=2\pi m$ for the initial
phase difference with $m=0,\pm 1, \pm 2,..$.

Taking the applied magnetic field along the $c$-axis (and normal to
the $CuO$ plane), that is ${\bf B}=(0,0,B)$, we obtain finally
\begin{equation}
M(B,\delta )=-M_0(\delta )\frac{b-b_{\mu }}{(1+b^2)(1+(b-b_{\mu
})^2)}
\end{equation}
for the chemically-induced sheet magnetization of the 2D Josephson
network.

Here $M_0(\delta )=J_0(\delta )/B_0$ with $J_0(\delta )$ defined
earlier (in what follows, $M_0(0)$ is $M_0(\delta \simeq 0)$),
$b=B/B_0$, and $b_{\mu }=B_{\mu }/B_0\simeq (k_BT\tau /\hbar )\delta
$ where $B_{\mu }(\delta )=(\mu _v\tau /\hbar )B_0$ is the
chemically-induced contribution (which disappears in optimally doped
systems with $\delta \simeq 0$), and $B_0=\Phi _0/wd$ is a
characteristic Josephson field.
\begin{figure*}[t]
\centerline{\includegraphics[width=9.0cm,clip=true]{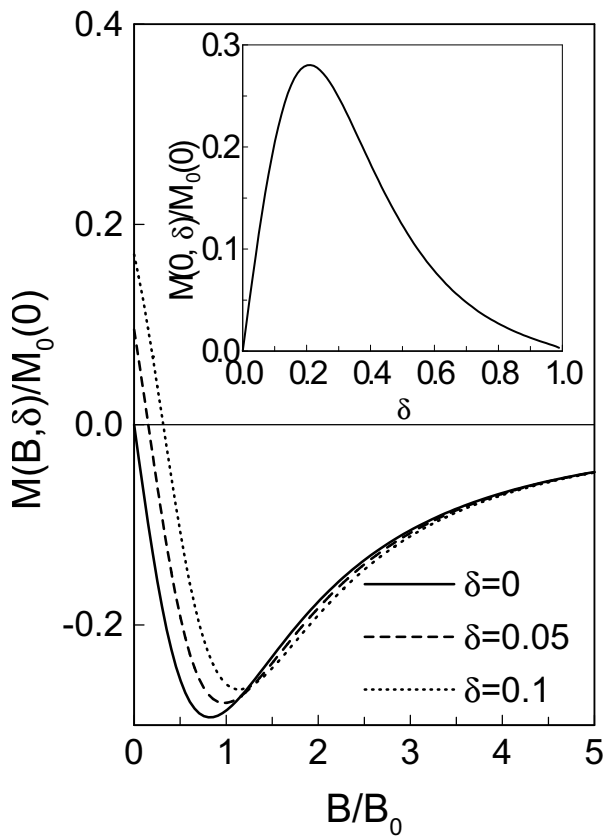}}
\caption{The magnetization $M(B,\delta )/M_0(0)$
 as a function of applied magnetic field $B/B_0$, according to
 Eq.(5), for different values of oxygen deficiency parameter:
 $\delta \simeq 0$ (solid line), $\delta =0.05$ (dashed line),
 and $\delta=0.1$ (dotted line). Inset: $\delta$ induced magnetization
 $M(0,\delta )/M_0(0)$ in a zero applied magnetic field (chemomagnetism).}
\end{figure*}
Fig.1 shows changes of the initial (stoichiometric) diamagnetic
magnetization $M/M_0$ (solid line) with oxygen deficiency
$\delta$. As is seen, even relatively small values of $\delta$
parameter render a low field Meissner phase strongly paramagnetic
(dotted and dashed lines). The inset of Fig.1 presents a true {\it
chemomagnetic} effect with concentration (deficiency) induced
magnetization $M(0,\delta )$ in zero magnetic field. According to
Eq.(5), the initially diamagnetic Meissner effect turns
paramagnetic as soon as the chemomagnetic contribution $B_{\mu
}(\delta )$ exceeds an applied magnetic field $B$. To see whether
this can actually happen in a real material, let us estimate a
magnitude of the chemomagnetic field $B_{\mu }$.
Typically~\cite{15,23}, for HTS single crystals $\lambda
_L(0)\approx 150nm$ and $d\simeq 10nm$, leading to $B_0\simeq
0.5T$. Using $\tau \simeq \hbar /\mu _v$ and $j_c=10^{10}A/m^2$ as
a pertinent characteristic time and the typical value of the
critical current density, respectively, we arrive at the following
estimate of the chemomagnetic field $B_{\mu }(\delta )\simeq
0.5B_0$ for $\delta =0.05$. Thus, the predicted chemically induced
PME should be observable for applied magnetic fields $B\simeq
0.5B_0\simeq 0.25T$ (which are actually much higher than the
fields needed to observe the previously discussed~\cite{21}
piezomagnetism and stress induced PME in high-$T_c$ ceramics).
Notice that for the above set of parameters, the Josephson length
$\lambda _J\simeq 1\mu m$, which means that the assumed in this
paper small-junction approximation (with $d\ll \lambda _J$) is
valid and the so-called "self-field" effects can be safely
neglected.
\begin{figure*}[t]
\centerline{\includegraphics[width=9.cm,clip=true]{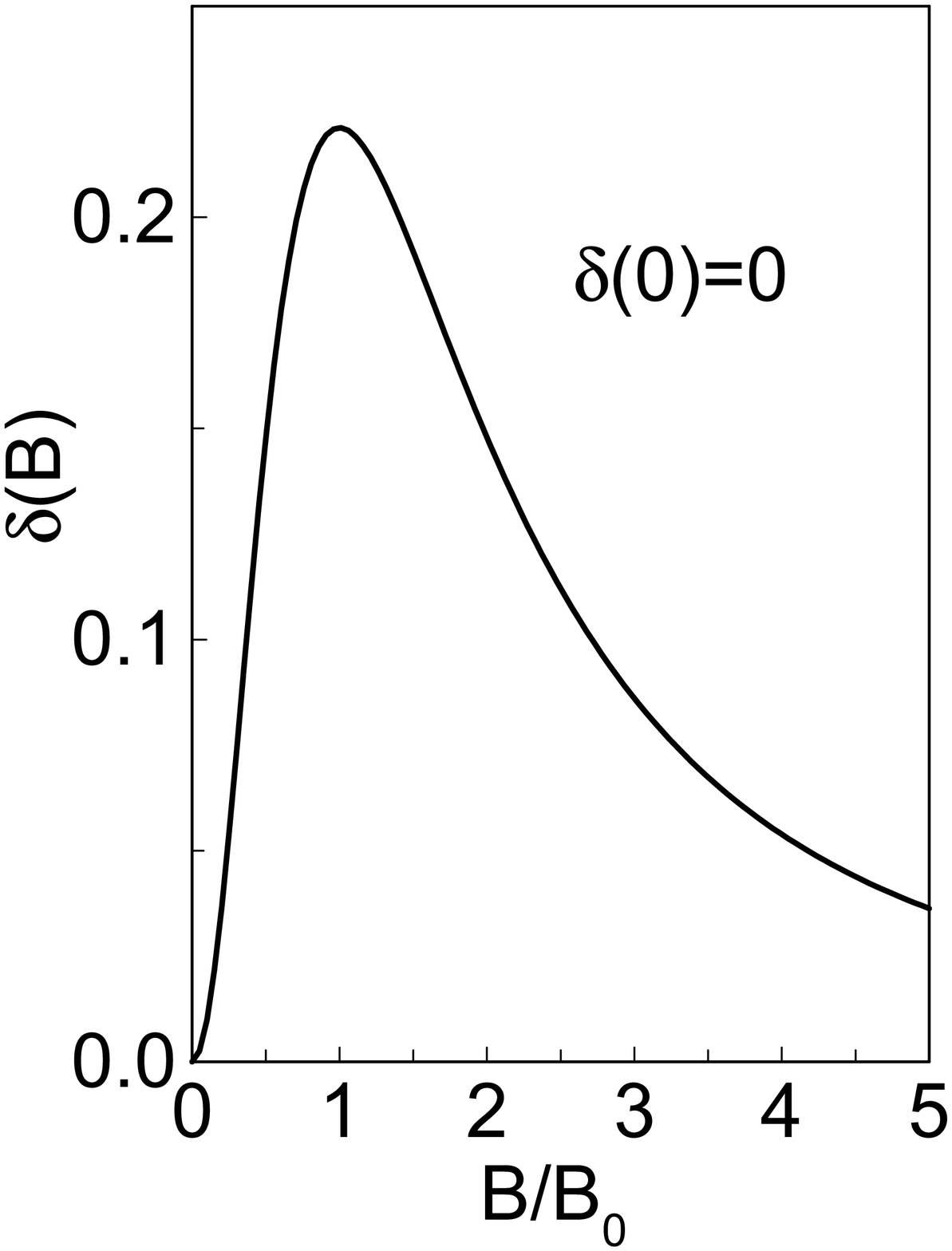}}
\caption{Magnetic field dependence of the oxygen
 deficiency parameter $\delta (B)$ (magnetoconcentration effect).}
\end{figure*}
So far, we neglected a possible field dependence of the chemical
potential $\mu _v$ of oxygen vacancies. However, in high enough
applied magnetic fields $B$, the field-induced change of the
chemical potential $\Delta \mu _v(B)\equiv \mu _v(B)-\mu _v(0)$
becomes tangible and should be taken into account. As is
well-known~\cite{28,29}, in a superconducting state $\Delta \mu
_v(B)=-M(B)B/n$, where $M(B)$ is the corresponding magnetization,
and $n$ is the relevant carriers number density. At the same time,
within our scenario, the chemical potential of a single oxygen
vacancy $\mu _v$ depends on the concentration of oxygen vacancies
(through deficiency parameter $\delta $). As a result, two
different effects are possible related respectively to magnetic
field dependence of $\mu _v(B)$ and to its dependence on
magnetization $\mu _v(M)$. The former is nothing else but a
superconducting analog of the so-called {\it magnetoconcentration}
effect (which was predicted and observed in inhomogeneously doped
semiconductors~\cite{30}) with field-induced creation of oxygen
vacancies $c_v(B)=c_v(0)\exp(-\Delta \mu _v(B)/k_BT)$, while the
latter (as we shall see in the next Section) results in a
"fishtail"-like behavior of the magnetization. Let us start with
the magnetoconcentration effect. Figure~2 depicts the predicted
field-induced creation of oxygen vacancies $\delta (B)=1-c_v(B)$
using the above-obtained magnetization $M(B,\delta )$ (see Fig.1
and Eq.(5)). We also assumed, for simplicity, a complete
stoichiometry of the system in a zero magnetic field (with $\delta
(0)=1-c_v(0)=0$). Notice that $\delta (B)$ exhibits a maximum at
$\delta _c\simeq 0.23$ for applied fields $B=B_0$ (in agreement
with the classical percolative behavior observed in
non-stoichiometric $YBa_2Cu_3O_{7-\delta }$
samples~\cite{15,16,24}). Finally, let us show that in underdoped
crystals the above-discussed osmotic mechanism of oxygen transport
is indeed much more effective than a traditional diffusion. Using
typical $YBCO$ parameters~\cite{23}, $\epsilon _0=0.01$, $\Omega
_0=a_0^3$ with $a_0=0.2nm$, and $K=115GPa$, we have $\mu
_v(0)=\epsilon _0K\Omega _0\simeq 1meV$ for
 a zero-field value of the chemical potential in
HTS crystals, which leads to creation of excess vacancies with
concentration $c_v(0)=e^{-\mu _v(0)/k_BT}\simeq 0.75$ (equivalent
to a deficiency value of $\delta (0)\simeq 0.25$) at $T=T_c$,
while the probability of oxygen diffusion in these materials
(governed by a rather high activation energy $U_d\simeq 1eV$) is
extremely slow under the same conditions because $D\propto
e^{-U_d/k_BT_c}\ll 1$. On the other hand, the change of the
chemical potential in applied magnetic field can reach as much
as~\cite{29} $\Delta \mu _v(B)\simeq 0.5meV$ for $B=0.5T$, which
is quite comparable with the above-mentioned zero-field value of
$\mu _v(0)$.
\begin{figure*}[t]
\centerline{\includegraphics[width=9.cm,clip=true]{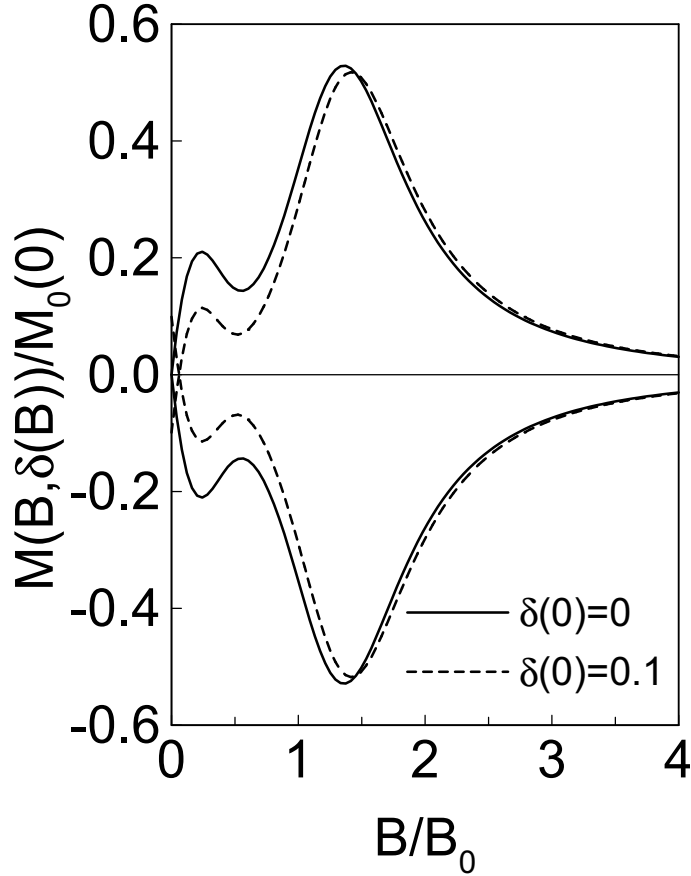}}
\caption{ A "fishtail"-like behavior of magnetization
 $m_f=M(B,\delta (B))/M_0(0)$ in applied magnetic field $B/B_0$ in the presence
 of magnetoconcentration effect (with field-induced oxygen vacancies
 $\delta (B)$, see Fig.2) for two values of field-free deficiency parameter:
 $\delta (0)\simeq 0$ (solid line), and $\delta
(0)=0.1$ (dashed line).}
\end{figure*}
Let us turn now to the second effect related to the magnetization
dependence of the chemical potential $\mu _v(M(B))$. In this case,
in view of Eq.(2), the phase difference will acquire an extra
$M(B)$ dependent contribution and as a result the r.h.s. of Eq.(5)
will become a nonlinear functional of $M(B)$. The numerical
solution of this implicit equation for the resulting magnetization
$m_{f}=M(B,\delta (B))/M_0(0)$ is shown in Fig.3 for the two
values of zero-field deficiency parameter $\delta (0)$. As is
clearly seen, $m_{f}$ exhibits a field-induced "fishtail"-like
behavior typical for underdoped crystals with intragrain
granularity (for symmetry and better visual effect we also plotted
$-m_{f}$ in the same figure). The extra extremum of the
magnetization appears when the applied magnetic field $B$ matches
an intrinsic chemomagnetic field $B_{\mu}(\delta (B))$ (which now
also depends on $B$ via the above-discussed magnetoconcentration
effect). Notice that a "fishtail" structure of $m_{f}$ manifests
itself even at zero values of field-free deficiency parameter
$\delta (0)$ (solid line in Fig.3) thus confirming a field-induced
nature of intrinsic granularity~\cite{14,15,22,23,24}. At the same
time, even a rather small deviation from the zero-field
stoichiometry (with $\delta (0)=0.1$) immediately brings about a
paramagnetic Meissner effect at low magnetic fields. Thus, the
present model predicts appearance of two interrelated phenomena,
Meissner paramagnetism at low fields and "fishtail" anomaly at
high fields. It would be very interesting to verify these
predictions experimentally in non-stoichiometric superconductors
with pronounced networks of planar defects.

\vspace{8mm} \leftline{\bf 3. MAGNETIC FIELD INDUCED POLARIZATION EFFECTS} \vspace{5mm}

In this Section, within the same model of JJAs created by a regular
2D network of twin-boundary (TB) dislocations with strain fields acting
as an insulating barrier between hole-rich domains in underdoped crystals,
we discuss charge-related effects which are actually dual to the above-described
phase-related chemomagnetic effects. More specifically, in what follows,
we are interested in the behavior of magnetic field induced electric polarization
(chemomagnetoelectricity) in chemically induced GS.

Recall that a conventional (zero-field) pair polarization operator
within the model under discussion reads~\cite{27,31}
\begin{equation}
{\bf p}=\sum_{i=1}^Nq_i {\bf x}_{i}
\end{equation}
In view of Eqs.(1), (2) and (6), and taking into account a usual
"phase-number" commutation relation, $[\phi _i,n_j]=i\delta
_{ij}$, it can be shown that the evolution of the pair polarization operator is
determined via the equation of motion
\begin{equation}
\frac{d{\bf p}}{dt}=\frac{1}{i\hbar}\left[ {\bf p},{\cal H}\right
] =\frac{2e}{\hbar }\sum_{ij}^NJ_{ij}\sin \phi _{ij}(t){\bf x}_{ij}
\end{equation}
Resolving the above equation, we arrive at the following net value
of the magnetic-field induced longitudinal (along $x$-axis)
electric polarization ${P}(\delta ,{\bf B}) \equiv <{p}_x(t)>$
and the corresponding effective junction charge
\begin{equation}
{Q}(\delta ,{\bf B})=\frac{2eJ_0} {\hbar \tau d} \int\limits_{0}^
{\tau }dt \int \limits_{0}^{t}dt'\int \frac{d^2x}{S}
\sin \phi ({\bf x}, t')xe^{-{\mid{{\bf x}}\mid}/d},
\end{equation}
where $S=2\pi d^2$ is properly defined normalization area, $\tau$
is a characteristic time (see Discussion), and we made a usual
substitution $\frac{1}{N}\sum_{ij}A_{ij}(t) \to \frac{1}{S}\int
d^2x A({\bf x},t)$ valid in the long-wavelength
approximation~\cite{27}.

To capture the very essence of the superconducting analog of the
chemomagnetoelectric effect, in what follows we assume for
simplicity that a {\it stoichiometric sample} (with $\delta \simeq
0$) does not possess any spontaneous polarization at zero magnetic
field, that is $P(0,0)=0$. According to Eq.(8), this condition
implies $\phi _{ij}^0=2\pi m$ for the initial phase difference
with $m=0,\pm 1, \pm 2,..$.

Taking the applied magnetic field along the $c$-axis (and normal to
the $CuO$ plane), that is ${\bf B}=(0,0,B)$, we obtain finally
\begin{equation}
Q(\delta ,B)=Q_0(\delta ) \frac{2{\tilde b}+b(1-{\tilde
b}^2)}{(1+b^2)(1+{\tilde b}^2)^2}
\end{equation}
for the magnetic field behavior of the effective junction charge in chemically
induced granular superconductors.
\begin{figure*}[t]
\centerline{\includegraphics[width=9cm,clip=true]{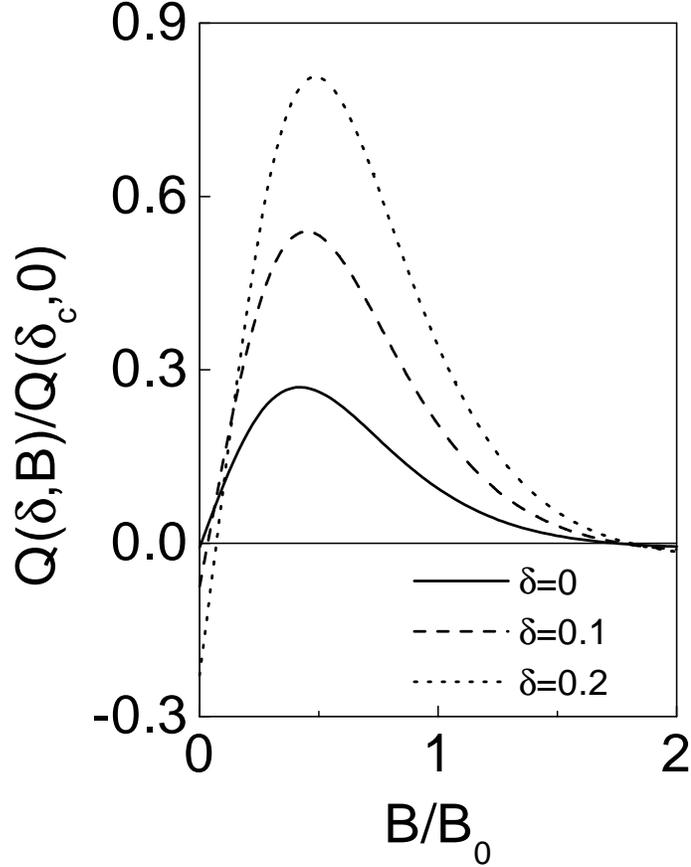}}
\caption{The effective junction charge $Q(\delta ,B)/Q(\delta
_c,0)$ (chemomagnetoelectric effect) as a function of applied
magnetic field $B/B_0$, according to
 Eq.(9), for different values of oxygen deficiency parameter: $\delta \simeq 0$ (solid line),
 $\delta =0.1$ (dashed line), and $\delta=0.2$ (dotted line).}
\end{figure*}
Here $Q_0(\delta )=e\tau J_0(\delta )/\hbar$ with $J_0(\delta )$
defined earlier, $b=B/B_0$, ${\tilde b}=b-b_{\mu }$, and $b_{\mu
}=B_{\mu }/B_0\simeq (k_BT\tau /\hbar )\delta $ where $B_{\mu
}(\delta )=(\mu _v\tau /\hbar )B_0$ is the chemically-induced
contribution (which disappears in optimally doped systems with
$\delta \simeq 0$), and $B_0=\Phi _0/wd$ is a characteristic
Josephson field.

Fig.4 shows changes of the initial (stoichiometric)
effective junction charge $Q/Q_0$ (solid line) with oxygen deficiency
$\delta$. Notice a sign change of $Q/Q_0$ (dotted and dashed lines)
driven by non-zero values of $\delta$ at low magnetic fields (a charge analog of
chemically induced PME). According to Eq.(9), the effective charge
changes its sign as soon as the chemomagnetic contribution $B_{\mu }(\delta )$
exceeds an applied magnetic field $B$.
\begin{figure*}[t]
\centerline{\includegraphics[width=9.cm,clip=true]{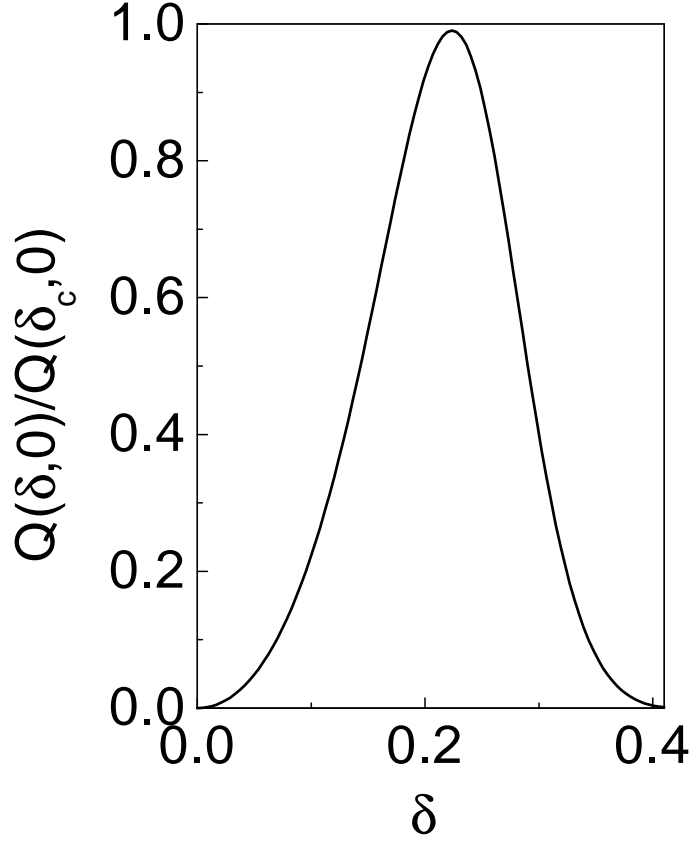}}
\caption{Chemically induced effective junction charge $Q(\delta
,0)/Q(\delta _c,0)$ in a zero applied magnetic field (true
chemoelectric effect).}
\end{figure*}
At the same time, Fig.5 presents a true {\it chemoelectric} effect
with concentration (deficiency) induced effective junction charge
$Q(\delta ,0)$ in zero magnetic field. Notice that $Q(\delta ,0)$
exhibits a maximum around $\delta _c\simeq 0.2$ (in agreement with
the classical percolative behavior observed in non-stoichiometric
$YBa_2Cu_3O_{7-\delta }$ samples~\cite{16}).

It is of interest also to consider the magnetic field behavior of
the concomitant effective flux capacitance $C\equiv \tau dQ(\delta ,B)/d\Phi $
which in view of Eq.(9) reads
\begin{equation}
C(\delta ,B)=C_0(\delta )\frac{1-3b{\tilde b}-3{\tilde
b}^2+b{\tilde b}^3}{(1+b^2)(1+{\tilde b}^2)^3},
\end{equation}
where $\Phi =SB$, and $C_0(\delta )=\tau Q_0(\delta )/\Phi _0$.

Fig.6 depicts the behavior of the effective flux capacitance
$C(\delta ,B)/C_0(0)$ in applied magnetic field for different
values of oxygen deficiency parameter: $\delta \simeq 0$ (solid
line), $\delta =0.1$ (dashed line), and $\delta=0.2$ (dotted
line). Notice a decrease of magnetocapacitance amplitude and its
peak shifting with increase of $\delta$ and sign change at low
magnetic fields which is another manifestation of the charge
analog of chemically induced PME (Cf. Fig.4).
\begin{figure*}[t]
\centerline{\includegraphics[width=9.cm,clip=true]{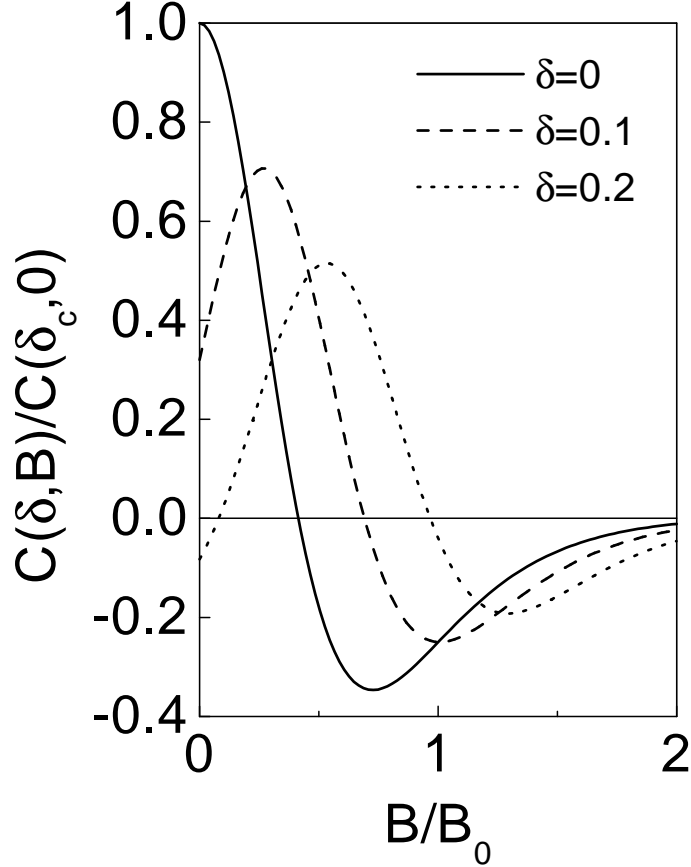}}
\caption{The effective flux capacitance $C(\delta ,B)/C(\delta
_c,0)$
 as a function of applied magnetic field $B/B_0$, according to Eq.(10),
 for different values of oxygen deficiency parameter: $\delta \simeq 0$
 (solid line), $\delta =0.1$ (dashed line), and $\delta=0.2$ (dotted line).}
\end{figure*}
Up to now, we neglected a possible field dependence of the
chemical potential $\mu _v$ of oxygen vacancies. Recall, however,
that in high enough applied magnetic fields $B$, the field-induced
change of the chemical potential $\Delta \mu _v(B)\equiv \mu
_v(B)-\mu _v(0)$ becomes tangible and should be taken into
account~\cite{28,29,32}. As a result, we end up with a
superconducting analog of the so-called {\it magnetoconcentration}
effect~\cite{32} with field induced creation of oxygen vacancies
$c_v(B)=c_v(0)\exp(-\Delta \mu _v(B)/k_BT)$ which in turn brings
about a "fishtail"-like behavior of the high-field
chemomagnetization (see Section 2 for more details).
\begin{figure*}[t]
\centerline{\includegraphics[width=9.cm,clip=true]{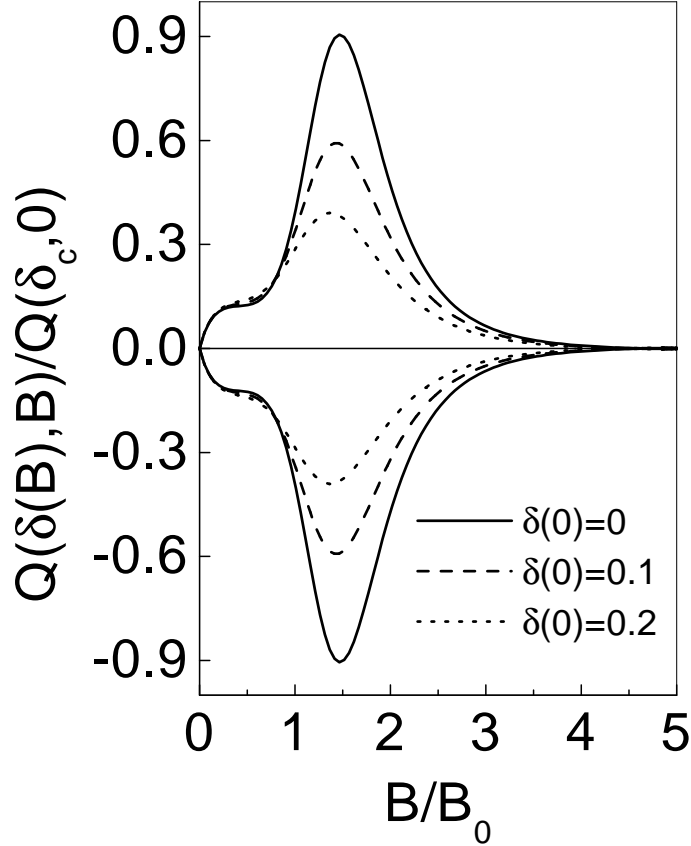}}
\caption{ A "fishtail"-like behavior of effective charge
 $Q(\delta (B),B)/Q(\delta
_c,0)$ in applied magnetic field $B/B_0$ in the presence
 of magnetoconcentration effect (with field-induced oxygen vacancies
 $\delta (B)$) for three values of field-free deficiency parameter:
 $\delta (0)\simeq 0$ (solid line), $\delta (0)=0.1$ (dashed line),
 and $\delta (0)=0.2$ (dotted line).}
\end{figure*}
Fig.7 shows the field behavior of the effective junction charge in
the presence of the above-mentioned magnetoconcentration effect.
As it is clearly seen, $Q(\delta (B),B)$ exhibits a
"fishtail"-like anomaly typical for previously discussed~\cite{32}
chemomagnetization in underdoped crystals with intragrain
granularity (for symmetry and better visual effect we also plotted
$-Q(\delta (B),B)$ in the same figure). This more complex
structure of the effective charge appears when the applied
magnetic field $B$ matches an intrinsic chemomagnetic field
$B_{\mu}(\delta (B))$ (which now also depends on $B$ via the
magnetoconcentration effect). Notice that a "fishtail" structure
of $Q(\delta (B),B)$ manifests itself even at zero values of
field-free deficiency parameter $\delta (0)$ (solid line in Fig.7)
thus confirming a field-induced nature of intrinsic granularity.
Likewise, Fig.8 depicts the evolution of the effective flux
capacitance $C(\delta (B),B)/C_0(0)$ in applied magnetic field
$B/B_0$ in the presence of magnetoconcentration effect (Cf.
Fig.6).
\begin{figure*}[t]
\centerline{\includegraphics[width=9.cm,clip=true]{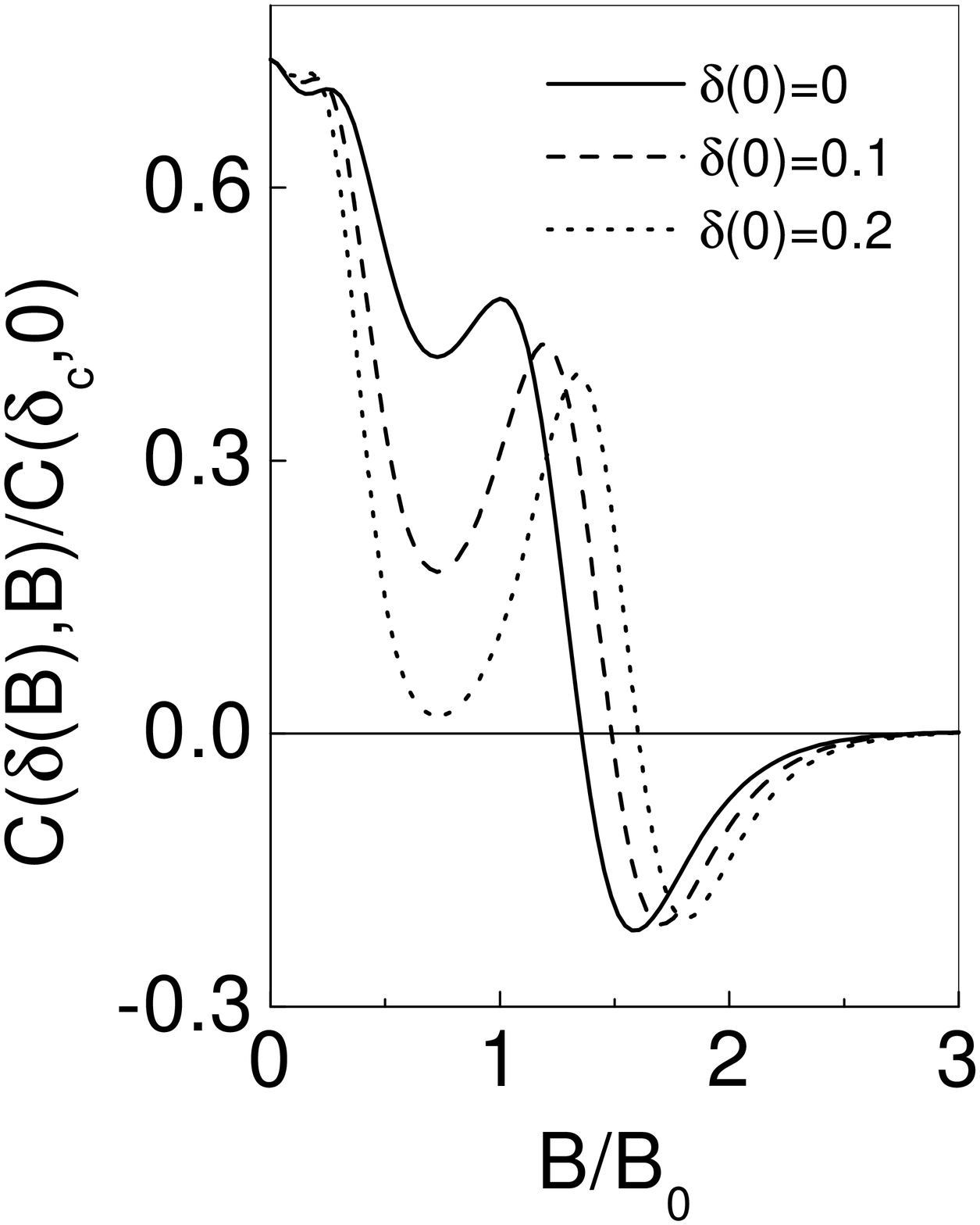}}
\caption{ The behavior of the effective flux capacitance
 $C(\delta (B),B)/C(\delta
_c,0)$ in applied magnetic field $B/B_0$ in the presence
 of magnetoconcentration effect for three values of field-free
 deficiency parameter: $\delta (0)\simeq 0$ (solid line), $\delta (0)=0.1$
 (dashed line), and $\delta (0)=0.2$ (dotted line).}
\end{figure*}
Thus, the present model predicts appearance of two interrelated
phenomena dual to the previously discussed behavior of
chemomagnetizm (see Section 2), namely a charge analog of Meissner
paramagnetism at low fields and a charge analog of "fishtail"
anomaly at high fields. To see whether these effects can be
actually observed in a real material, let us estimate an order of
magnitude of the main model parameters.

Using typical for HTS single crystals values of $\lambda _L(0)
\simeq 150nm$, $d \simeq 10nm$, and $j_c \simeq 10^{10}A/m^2$, we
arrive at the following estimates of the characteristic $B_0
\simeq 0.5T$  and chemomagnetic $B_{\mu }(\delta ) \simeq 0.5B_0$
fields, respectively. So, the predicted charge analog of PME
should be observable for applied magnetic fields $B < 0.25T$.
Notice that, for the above set of parameters, the Josephson length
is of the order of $\lambda _J \simeq 1\mu m$, which means that
the assumed in this paper small-junction approximation is valid
and the "self-field" effects can be safely neglected.

Furthermore, the characteristic frequencies $\omega \simeq \tau
^{-1}$ needed to probe the suggested here effects are related to
the processes governed by tunneling relaxation times $\tau \simeq
\hbar /J_0(\delta )$. Since for oxygen deficiency parameter
$\delta =0.1$ the chemically-induced zero-temperature Josephson
energy in non-stoichiometric $YBCO$ single crystals is of the
order of $J_0(\delta ) \simeq k_B T_C \delta \simeq 1meV$, we
arrive at the required frequencies of $\omega \simeq 10^{13}Hz$
and at the following estimates of the effective junction charge
$Q_0 \simeq e=1.6\times 10^{-19}C$ and flux capacitance $C_0
\simeq 10^{-18}F$. Notice that the above estimates fall into the
range of parameters used in typical experiments for studying the
single-electron tunneling effects both in JJs and
JJAs~\cite{1,2,13,33} suggesting thus quite an optimistic
possibility to observe the above-predicted field induced effects
experimentally in non-stoichiometric superconductors with
pronounced networks of planar defects or in artificially prepared
JJAs. (It is worth mentioning that a somewhat similar behavior of
the magnetic field induced charge and related flux capacitance has
been observed in 2D electron systems~\cite{34}.)

And finally, it can be easily verified that, in view of
Eqs.(6)-(8), the field-induced Coulomb energy of the
oxygen-depleted region within our model is given by
\begin{equation}
E_C(\delta ,B)\equiv \left < \sum_{ij}^N \frac{q_iq_j}{2C_{ij}}
\right >=\frac{Q^2(\delta ,B)}{2C(\delta ,B)}
\end{equation}
with $Q(\delta ,B)$ and $C(\delta ,B)$ defined by Eqs. (9) and
(10), respectively.

A thorough analysis of the above expression reveals that in the
PME state (when $B\ll B_{\mu }$) the chemically-induced granular
superconductor is in the so-called Coulomb blockade regime (with
$E_C>J_0$), while in the ''fishtail'' state (for $B\ge B_{\mu }$)
the energy balance tips in favor of tunneling (with $E_C<J_0$). In
particular, we obtain that $E_C(\delta ,B=0.1B_{\mu
})=\frac{\pi}{2}J_0(\delta )$ and $E_C(\delta ,B=B_{\mu
})=\frac{\pi}{8}J_0(\delta )$. It would be also interesting to
check this phenomenon of field-induced weakening of the Coulomb
blockade experimentally.

\vspace{8mm} \leftline{\bf 4. CONCLUSION} \vspace{5mm}

In conclusion, within a realistic model of 2D Josephson junction arrays (created by
2D network of twin boundary dislocations with strain fields acting as
an insulating barrier between hole-rich domains in underdoped
crystals), a few novel effects expected to occur in intrinsically
granular material are predicted, including phase-related and (dual) charge-related
phenomena. The conditions under which these effects can be experimentally measured in
non-stoichiometric high-$T_c$ superconductors were discussed.

\vspace{12mm} \leftline{\bf ACKNOWLEDGMENTS} \vspace{5mm}

This work was done during my stay at the Center for Physics of
Fundamental Interactions (Instituto Superior T\'ecnico, Lisbon)
and was partially funded by the FCT. I thank Pedro Sacramento and
Vitor Vieira for hospitality and interesting discussions on the
subject. I am also indebted to Anant Narlikar for his invitation
to make this contribution for the Special Golden Jubilee volume of
the Studies.

\vspace{16mm}

\vspace{30mm}

\begin{thebibliography}{99}
\bibitem{1} M. Iansity, A.J. Johnson, C.J. Lobb et al., Phys. Rev. Lett. {\bf 60} (1988) 2414\vspace*{-0.25cm}
\bibitem{2} D.B. Haviland, L.S. Kuzmin, P. Delsing et al., Z. Phys. B {\bf 85} (1991) 339\vspace*{-0.25cm}
\bibitem{3} H.S.J. van der Zant, Physica B {\bf 222} (1996) 344\vspace*{-0.25cm}
\bibitem{4} V.V. Ryazanov, V.A. Oboznov, A.Yu. Rusanov et al., Phys. Rev.
Lett. {\bf 86} (2001) 2427\vspace*{-0.25cm}
\bibitem{5} A.A. Golubov, M.Yu. Kupriyanov, and Ya.V. Fominov, JETP Lett. {\bf 75} (2002) 588\vspace*{-0.25cm}
\bibitem{6} P.M. Ostrovsky and M.V. Feigel'man, JETP Lett. {\bf 79} (2004) 489\vspace*{-0.25cm}
\bibitem{7} F.M. Araujo-Moreira, P. Barbara, A.B. Cawthorne et al., Phys. Rev. Lett. {\bf 78} (1997) 4625;
P. Barbara, F.M. Araujo-Moreira, A.B. Cawthorne et al., Phys. Rev.
B {\bf 60} (1999) 7489; F.M. Araujo-Moreira, W. Maluf, and S.
Sergeenkov, Eur. Phys. J. B {\bf 44} (2005) 33\vspace*{-0.25cm}
\bibitem{8} S. Sergeenkov and F.M. Araujo-Moreira, JETP Lett. {\bf 80} (2004) 580\vspace*{-0.25cm}
\bibitem{9} I.V. Krive, S.I. Kulinich, R.I. Shekhter et al., Low Temp. Phys. {\bf 30} (2004) 554\vspace*{-0.25cm}
\bibitem{10} L.B. Ioffe, M.V. Feigel'man, A. Ioselevich et al., Nature
{\bf 415} (2002) 503\vspace*{-0.25cm}
\bibitem{11} D. Born, V.I. Shnyrkov, W. Krechet et al., Phys. Rev. B {\bf 70} (2004) 180501\vspace*{-0.25cm}
\bibitem{12} A.B. Zorin, JETP {\bf 98} (2004) 1250\vspace*{-0.25cm}
\bibitem{13}  Yu. Makhlin, G. Sch\"on, and A. Shnirman, Rev. Mod. Phys. {\bf 73} (2001) 357\vspace*{-0.25cm}
\bibitem{14} K.M. Lang, V. Madhavan, J.E. Hoffman et al., Nature {\bf 415} (2002)
 412\vspace*{-0.25cm}
\bibitem{15} M. Daeumling, J.M. Seuntjens, D.C. Larbalestier et al., Nature
{\bf 346} (1990) 332; I.M. Babich and G.P. Mikitik, JETP Lett. {\bf
64} (1996) 586\vspace*{-0.25cm}
\bibitem{16} V.F. Gantmakher, A.M. Neminskii, and D.V. Shovkun,
JETP Lett. {\bf 52} (1990) 630\vspace*{-0.25cm}
\bibitem{17} S. Sergeenkov, {\em Studies of High Temperature Superconductors}, {\bf v. 39}
 (Ed. A. Narlikar), Nova Sci. Publishers, NY (2001), pp. 117-131;
F.M. Araujo-Moreira, P. Barbara, A.B. Cawthorne, and C.J. Lobb,
{\em Studies of High Temperature Superconductors}, {\bf v. 43}
(Ed. A. Narlikar), Nova Sci. Publishers, NY (2002), pp. 227-242;
S. Sergeenkov, {\em New Developments in Superconductivity
Research} (Ed. R.W. Stevens), Nova Sci. Publishers, NY (2003), pp.
18-45\vspace*{-0.25cm}
\bibitem{18} V. Kataev, N. Knauf, W. Braunisch et al.,
JETP Lett. {\bf 58} (1993) 636; A.K. Geim, S.V. Dubonos, J.G.S. Lok
et al., Nature {\bf 396} (1998) 144; M.S. Li, Phys. Rep. {\bf 376} (2003) 133\vspace*{-0.25cm}
\bibitem{19} C. De Leo and G. Rotoli, Phys. Rev. Lett. {\bf 89} (2002) 167001\vspace*{-0.25cm}
\bibitem{20} I. N. Khlyustikov and M. S. Khaikin, JETP {\bf 48} (1978) 583; M. S. Khaikin and I. N. Khlyustikov,
JETP Lett. {\bf 33} (1981) 158\vspace*{-0.25cm}
\bibitem{21} S. Sergeenkov, JETP Lett. {\bf 70} (1999) 36\vspace*{-0.25cm}
\bibitem{22} G.Yang, P. Shang, S.D. Sutton et al., Phys.Rev.B {\bf 48} (1993) 4054\vspace*{-0.25cm}
 \bibitem{23} A. Gurevich and E.A. Pashitskii, Phys.Rev.B {\bf 56} (1997) 6213\vspace*{-0.25cm}
 \bibitem{24} B.H. Moeckley, D.K. Lathrop, and R.A. Buhrman, Phys. Rev. B {\bf 47} (1993) 400\vspace*{-0.25cm}
\bibitem{25} L.A. Girifalco, {\it Statistical Physics of Materials} (A
Wiley-Interscience Publication, New York, 1973)\vspace*{-0.25cm}
\bibitem{26} S. Sergeenkov, J. Appl. Phys. {\bf 78} (1995) 1114\vspace*{-0.25cm}
\bibitem{27} S. Sergeenkov, JETP Lett. {\bf 76} (2002) 170\vspace*{-0.25cm}
\bibitem{28} A.A. Abrikosov, {\it Fundamentals of the Theory of Metals}
(Elsevier, Amsterdam, 1988)\vspace*{-0.25cm}
\bibitem{29} S. Sergeenkov and M. Ausloos, JETP {\bf 89} (1999) 140\vspace*{-0.25cm}
\bibitem{30} A.A. Akopyan, S.S. Bolgov, A.P. Savchenko et al.,
Sov. Phys. Semicond. {\bf 24} (1990) 1167\vspace*{-0.25cm}
\bibitem{31} S. Sergeenkov, J. de Physique I (France) {\bf 7} (1997) 1175\vspace*{-0.25cm}
\bibitem{32} S. Sergeenkov, JETP Lett. {\bf 77} (2003) 94\vspace*{-0.25cm}
\bibitem{33} P.J.M. van Bentum, H. van Kempen, L.E.C. van de Leemput et al., Phys. Rev. Lett. {\bf 60} (1988) 369\vspace*{-0.25cm}
\bibitem{34} W. Chen, T.P. Smith, M. Buttiker et al., Phys. Rev. Lett. {\bf 73} (1994) 146\vspace*{-0.25cm}
\end{thebibliography}
\end{document}